\documentclass{article}
\usepackage{makeidx}
\usepackage{graphicx}
\usepackage{amsmath}
\usepackage{amsfonts}
\usepackage{amssymb}

\begin{document}

\title{Quantum Cooperative Games}
\author{A. Iqbal and A.H. Toor\\Electronics Department, Quaid-i-Azam University, \\Islamabad, Pakistan\\email: el1aiqbal@qau.edu.pk\\}
\maketitle
\begin{abstract}
We study two forms of a symmetric cooperative game played by three players,
one classical and other quantum. In its classical form making a coalition
gives advantage to players and they are motivated to do so. However in its
quantum form the advantage is lost and players are left with no motivation to
make a coalition.
\end{abstract}

\section{Introduction}

Many situations in the recent research in quantum games \cite{meyer, Eisert,
marinatto} appear to be based on a general idea that is quite interesting as
well. It is to take a classical game exhibiting certain features, generalize
it to quantum domain, and see how the situation changes in the course of this
generalization. In this course noncooperative games have attracted an earlier
attention with the ruling solution concept of a Nash equilibrium (NE). This
development looks reasonable because in classical game theory as well the
earlier research was focused on noncooperative games and interest in coalition
formation was revived later. Players in noncooperative games are not able to
form binding agreements even if they may communicate. On the other hand the
distinguishing feature of cooperative games is a strong incentive to work
together to receive the largest total payoff. These games allow players to
form coalitions, binding agreements, pay compensations, make side payments
etc. In fact, von Neumann and Morgenstern \cite{neumann} in their pioneering
work in the theory of games offered models of coalition formation where the
strategy of each player consists of choosing the coalition he wishes to join.
In coalition games, that are part of cooperative game theory, the
possibilities of the players are described by the available resources of
different groups (coalitions) of players. Joining a group or remaining outside
is part of strategy of a player affecting his/her payoff. Recent work in
quantum games \cite{meyer, Eisert, marinatto} gives rise to a natural and
interesting question: what is the possible quantum mechanical role in
cooperative games that are an important part of the classical game theory? In
our opinion it may be quite interesting, and fruitful as well, to investigate
coalitions in quantum versions of cooperative games. Our motivation in present
paper is to investigate what might happen to the advantage of forming a
coalition in a quantum game compared to its classical analogue. We rely on the
concepts and ideas of von Neumann's cooperative game theory \cite{neumann} and
consider a three-player coalition game in a quantum form. We then compare it
to the classical version of the game and see how the advantage of forming a
coalition can be affected.

In usual classical analysis of the coalition games the notion of a strategy
disappears; the main features are those of a coalition and the value or worth
of the coalition. The underlying assumption is that each coalition can
guarantee its members a certain amount called the ``value of a coalition''
\cite{burger}\textit{.} \textit{The value of coalition measures the worth the
coalition possesses and is characterized as the payoff which the coalition can
assure for itself by selecting an appropriate strategy, whereas the `odd man'
can prevent the coalition from getting more than this amount}. Using this idea
we study cooperative games in quantum settings to see how advantages of making
coalitions can be influenced in the new settings. The preferable scheme to us
to play a quantum game has been recently proposed by Marinatto and Weber
\cite{marinatto}. In this scheme an initial quantum state is prepared by an
arbiter and forwarded to the players. Each player possesses two quantum
unitary and Hermitian operators i.e. the identity $I$ and the inversion or
Pauli spin-flip operator $\sigma$. Players apply the operators with classical
probabilities on the initial quantum state and send the quantum state to the
`measuring agent' who decides the payoffs the players should get. Interesting
feature in this scheme is that the classical game is reproduced when the
initial quantum state becomes unentangled \cite{marinatto}. Classical game is
therefore embedded in the quantum version of the game.

In this paper using Marinatto and Weber's scheme \cite{marinatto} we find a
quantum form of a symmetric cooperative game played by three players. In
classical form of this game any two players out of three get an advantage when
they successfully form a coalition and play the same strategy. We find a
quantum form of this game where the advantage for coalition forming is lost
and players are left with no motivation to cooperate.

\section{A three player symmetric cooperative game}

\subsection{Classical form}

A classical three person normal form game \cite{burger} is given by three
non-empty sets $\Sigma_{A}$, $\Sigma_{B}$, and $\Sigma_{C}$, the strategy sets
of the players $A$,$B$, and $C$ and three real valued functions $P_{A}$,
$P_{B}$, and $P_{C}$ defined on $\Sigma_{A}\times\Sigma_{B}\times\Sigma_{C}$.
The product space $\Sigma_{A}\times\Sigma_{B}\times\Sigma_{C}$ is the set of
all tuples $(\sigma_{A},\sigma_{B},\sigma_{C})$ with $\sigma_{A}\in\Sigma_{A}%
$, $\sigma_{B}\in\Sigma_{B}$ and $\sigma_{C}\in\Sigma_{C}$. A strategy is
understood as such a tuple $(\sigma_{A},\sigma_{B},\sigma_{C})$ and $P_{A}$,
$P_{B}$, $P_{C}$ are payoff functions of the three players. The game is
usually denoted as $\Gamma=\left\{  \Sigma_{A},\Sigma_{B},\Sigma_{C}%
;P_{A},P_{B},P_{C}\right\}  $. Let $\Re=\left\{  A,B,C\right\}  $ be the set
of players and $\wp$ be an arbitrary subset of $\Re$. The players in $\wp$ may
form a coalition so that, for all practical purposes, the coalition $\wp$
appears as a single player. It is expected that players in $\Re-\wp$ will form
an opposing coalition and the game has two opposing ``coalition players'' i.e.
$\wp$ and $\Re-\wp$.

We study quantum version of an example of a classical three player cooperative
game discussed in ref. \cite{burger} . Each of three players $A,B $ and $C$
chooses one of the two strategies $1,2$. If the three players choose the same
strategy there is no payoff; otherwise, the two players who have chosen the
same strategy receive one unit of money each from the `odd man.' Payoff
functions $P_{A}$, $P_{B}$ and $P_{C}$ for players $A,B$ and $C$ respectively
are given as \cite{burger}%

\begin{align}
P_{A}(1,1,1)  &  =P_{A}(2,2,2)=0\nonumber\\
P_{A}(1,1,2)  &  =P_{A}(2,2,1)=P_{A}(1,2,1)=P_{A}(2,1,2)=1\nonumber\\
P_{A}(1,2,2)  &  =P_{A}(2,1,1)=-2 \label{payoffs}%
\end{align}
with similar expressions for $P_{B}$ and $P_{C}$. Suppose $\wp=\left\{
B,C\right\}  $; hence $\Re-\wp=\left\{  A\right\}  $. The coalition game
represented by $\Gamma_{\wp}$ is given by the following payoff matrix \cite{burger}%

\begin{equation}%
\begin{tabular}
[c]{lllll}%
&  & $\left[  1\right]  $ &  & $\left[  2\right]  $\\
$\left[  11\right]  $ &  & $0$ &  & $2$\\
$\left[  12\right]  $ &  & $-1$ &  & $-1$\\
$\left[  21\right]  $ &  & $-1$ &  & $-1$\\
$\left[  22\right]  $ &  & $2$ &  & $0$%
\end{tabular}
\end{equation}
Here the strategies $\left[  12\right]  $ and $\left[  21\right]  $ are
dominated by $\left[  11\right]  $ and $\left[  22\right]  $. After
eliminating these dominated strategies the payoff matrix becomes%

\begin{equation}%
\begin{tabular}
[c]{lll}%
& $\left[  1\right]  $ & $\left[  2\right]  $\\
$\left[  11\right]  $ & $0$ & $2$\\
$\left[  22\right]  $ & $2$ & $0$%
\end{tabular}
\end{equation}
It is seen that the mixed strategies%

\begin{align}
&  \frac{1}{2}\left[  11\right]  +\frac{1}{2}\left[  22\right] \label{cltC}\\
&  \frac{1}{2}\left[  1\right]  +\frac{1}{2}\left[  2\right]  \label{lftC}%
\end{align}
are optimal for $\wp$ and $\Re-\wp$ respectively. With these strategies a
payoff $1$ for players $\wp$ is assured for all strategies of the opponent;
hence, the value of the coalition $\upsilon(\Gamma_{\wp})$ is $1$ i.e.
$\upsilon(\left\{  B,C\right\}  )=1$. Since $\Gamma$ is a zero-sum game
$\upsilon(\Gamma_{\wp})$ can also be used to find $\upsilon(\Gamma_{\Re-\wp})$
as $\upsilon(\left\{  A\right\}  )=-1$. The game is also symmetric and one can write%

\begin{align}
\upsilon(\Gamma_{\wp})  &  =1\text{, \ \ and\ \ \ }\upsilon(\Gamma_{\Re-\wp
})=-1\text{ or}\nonumber\\
\upsilon(\left\{  A\right\}  )  &  =\upsilon(\left\{  B\right\}
)=\upsilon(\left\{  C\right\}  )=-1\nonumber\\
\upsilon(\left\{  A,B\right\}  )  &  =\upsilon(\left\{  B,C\right\}
)=\upsilon(\left\{  C,A\right\}  )=1 \label{VcltC}%
\end{align}

\subsection{Quantum form}

In quantum form of this three player game the players implement their
strategies by applying the identity operators in their possession with
probabilities $p,q,$ and $r$ respectively on the initial quantum state. In
Marinatto and Weber's scheme \cite{marinatto} the Pauli spin-flip or simply
the inversion operator $\sigma$ is then applied with probabilities
$(1-p),(1-q),$ and $(1-r)$ by players $A$, $B$ and $C$ respectively. If
$\rho_{in}$ is the density matrix corresponding to initial quantum state the
final state after players have played their strategies corresponds to \cite{iqbal}%

\begin{equation}
\rho_{fin}=\underset{U=I,\sigma}{\sum}\Pr(U_{A})\Pr(U_{B})\Pr(U_{C}%
)U_{A}\otimes U_{B}\otimes U_{C}\rho_{in}U_{A}^{\dagger}\otimes U_{B}%
^{\dagger}\otimes U_{C}^{\dagger} \label{finstat}%
\end{equation}
where the unitary and Hermitian operator $U$ can be either $I$ or $\sigma$.
$\Pr(U_{A})$, $\Pr(U_{B})$ and $\Pr(U_{C})$ are the probabilities with which
players $A$, $B$ and $C$ apply the operator $U$ on the initial state
respectively. $\rho_{fin}$ corresponds to a convex combination of all possible
quantum operations. Let the arbiter prepares the following three qubit pure
initial quantum state%

\begin{equation}
\left|  \psi_{in}\right\rangle =\underset{i,j,k=1,2}{\sum}c_{ijk}\left|
ijk\right\rangle \text{, \ \ where \ \ }\underset{i,j,k=1,2}{\sum}\left|
c_{ijk}\right|  ^{2}=1 \label{instate}%
\end{equation}
where the eight basis vectors of this quantum state are $\left|
ijk\right\rangle $ for $i,j,k=1,2$. The initial state (\ref{instate}) can be
imagined as a global state (in a $2\otimes2\otimes2$ dimensional Hilbert
space) of three two-state quantum systems or `qubits'. A player applies the
unitary operators $I$ and $\sigma$ with classical probabilities on $\rho_{in}$
during his `move' or `strategy' operation. Fig. 1 shows the scheme to play
this three player quantum game where players $B$ and $C$ form a coalition and
player $A$ is `leftout'.%

\begin{figure}
[ptb]
\begin{center}
\phantom{\rule{3.4255in}{2.7994in}}\caption{A three player quantum game played
with Marinatto and Weber's scheme. Players $B$ and $C$ form a coalition. $I$
is the identity and $\sigma$ is the inversion operator.}%
\label{Fig. 1}%
\end{center}
\end{figure}

Let the matrix of three player game be given by $24$ constants $\alpha
_{t},\beta_{t},\gamma_{t}$ with $1\leq t\leq8$ \cite{iqbal}. We write the
payoff operators for players $A,B,$ and $C$ as \cite{marinatto}%

\begin{align}
(P_{A,B,C})_{oper}  &  =\alpha_{1},\beta_{1},\gamma_{1}\left|
111\right\rangle \left\langle 111\right|  +\alpha_{2},\beta_{2},\gamma
_{2}\left|  211\right\rangle \left\langle 211\right|  +\nonumber\\
&  \alpha_{3},\beta_{3},\gamma_{3}\left|  121\right\rangle \left\langle
121\right|  +\alpha_{4},\beta_{4},\gamma_{4}\left|  112\right\rangle
\left\langle 112\right|  +\nonumber\\
&  \alpha_{5},\beta_{5},\gamma_{5}\left|  122\right\rangle \left\langle
122\right|  +\alpha_{6},\beta_{6},\gamma_{6}\left|  212\right\rangle
\left\langle 212\right|  +\nonumber\\
&  \alpha_{7},\beta_{7},\gamma_{7}\left|  221\right\rangle \left\langle
221\right|  +\alpha_{8},\beta_{8},\gamma_{8}\left|  222\right\rangle
\left\langle 222\right|  \label{payoper}%
\end{align}
Payoffs to players $A,B,$ and $C$\ are then obtained as mean values of these
operators \cite{marinatto}\ %

\begin{equation}
P_{A,B,C}(p,q,r)=Trace\left[  (P_{A,B,C})_{oper}\rho_{fin}\right]
\end{equation}
where, for convenience, we identify the players' moves only by the numbers
$p,q$ and $r$. The cooperative game of eq. (\ref{payoffs}) with the classical
payoff functions $P_{A}$, $P_{B}$ and $P_{C}$ for players $A$ $B$ and $C$
respectively, together with the definition of payoff operators for these
players in eq. (\ref{payoper}), imply that%

\begin{equation}
\alpha_{1}=\alpha_{8}=0\text{, \ \ \ \ }\alpha_{3}=\alpha_{4}=\alpha
_{6}=\alpha_{7}=1\text{ \ \ and\ \ \ }\alpha_{2}=\alpha_{5}=-2
\end{equation}
With these constants the payoff to player $A$, for example, can be found as%

\begin{equation}
P_{A}(p,q,r)=\left[
\begin{array}
[c]{c}%
-4rq-2p+2pr+2pq+r+q\\
-4rq+2p-2pr-2pq+3r+3q-2\\
4rq+2pr-2pq-3r-q+1\\
4rq-2pr+2pq-r-3q+1
\end{array}
\right]  \left[
\begin{array}
[c]{c}%
\left|  c_{111}\right|  ^{2}+\left|  c_{222}\right|  ^{2}\\
\left|  c_{211}\right|  ^{2}+\left|  c_{122}\right|  ^{2}\\
\left|  c_{121}\right|  ^{2}+\left|  c_{212}\right|  ^{2}\\
\left|  c_{112}\right|  ^{2}+\left|  c_{221}\right|  ^{2}%
\end{array}
\right]  \label{Poff}%
\end{equation}
Similarly payoffs to players $B$ and $C$ can be obtained. Classical mixed
strategy payoffs can be recovered from the eq. (\ref{Poff}) by taking $\left|
c_{111}\right|  ^{2}=1$. The classical game is therefore imbedded in its
quantum form.

The classical form of this game is symmetric in the sense that payoff to a
player depends on his/her strategy and not on his/her identity. These
requirements making symmetric the three-player game are written as%

\begin{align}
P_{A}(p,q,r)  &  =P_{A}(p,r,q)=P_{B}(q,p,r)=P_{B}(r,p,q)\nonumber\\
&  =P_{C}(r,q,p)=P_{A}(q,r,p) \label{rqmnts}%
\end{align}
Now in this quantum form of the game $P_{A}(p,q,r)$ becomes same as
$P_{A}(p,r,q)$ when%

\begin{equation}
\left|  c_{121}\right|  ^{2}+\left|  c_{212}\right|  ^{2}=0\text{,
\ \ \ \ \ }\left|  c_{112}\right|  ^{2}+\left|  c_{221}\right|  ^{2}=0
\label{rqmnts1}%
\end{equation}
and then payoff to a $p$ player remains same when other two players
interchange their strategies. The symmetry conditions (\ref{rqmnts}) hold if,
together with eqs. (\ref{rqmnts1}), following relations are also true%

\begin{equation}%
\begin{array}
[c]{cc}%
\alpha_{1}=\beta_{1}=\gamma_{1}, & \alpha_{5}=\beta_{6}=\gamma_{7}\\
\alpha_{2}=\beta_{3}=\gamma_{4}, & \alpha_{6}=\beta_{5}=\gamma_{6}\\
\alpha_{3}=\beta_{2}=\gamma_{3}, & \alpha_{7}=\beta_{7}=\gamma_{5}\\
\alpha_{4}=\beta_{4}=\gamma_{2}, & \alpha_{8}=\beta_{8}=\gamma_{8}%
\end{array}
\end{equation}
These form the extra restrictions on the constants of payoff matrix and,
together with the conditions (\ref{rqmnts1}), give a three player symmetric
game in a quantum form. No subscript in a payoff expression is then needed and
$P(p,q,r)$ represents the payoff to a $p$ player against two other players
playing $q$ and $r$. The payoff $P(p,q,r)$ is found as%

\begin{align}
P(p,q,r)  &  =(\left|  c_{111}\right|  ^{2}+\left|  c_{222}\right|
^{2})(-4rq-2p+2pr+2pq+r+q)+\nonumber\\
&  (\left|  c_{211}\right|  ^{2}+\left|  c_{122}\right|  ^{2}%
)(-4rq+2p-2pr-2pq+3r+3q-2)\nonumber\\
&  \label{Qpayoff}%
\end{align}
The term `mixed strategy' in the quantum form of this game is defined as being
a convex combination of quantum strategies with classical probabilities. For
this assume that the pure strategies $[1]$ and $[2]$ correspond to $p=0$ and
$p=1$ respectively. The mixed strategy $n\left[  1\right]  +(1-n)\left[
2\right]  $, where $0\leq n\leq1$, means that the strategy $\left[  1\right]
$ is played with probability $n$ and $\left[  2\right]  $ with probability
$(1-n)$. Now suppose the coalition $\wp$ plays the following mixed strategy%

\begin{equation}
l[11]+(1-l)[22] \label{cltQ}%
\end{equation}
where the strategy $[11]$ means that both players in the coalition $\wp$ apply
the identity operator $I$ with zero probability. Similarly the strategy $[22]$
can be defined. The strategy of the coalition in eq. (\ref{cltQ}) means that
the coalition $\wp$ plays $[11]$ with probability $l$ and $[22]$ with
probability $1-l$. Similarly we suppose the player in $\Re-\wp$ plays
following mixed strategy%

\begin{equation}
m[1]+(1-m)[2] \label{lftQ}%
\end{equation}
In this case the payoff to the coalition $\wp$ is obtained as%

\begin{align}
P_{\wp}  &  =(lm)P_{\wp\lbrack111]}+l(1-m)P_{\wp\lbrack112]}+\nonumber\\
&  (1-l)mP_{\wp\lbrack221]}+(1-l)(1-m)P_{\wp\lbrack222]} \label{cltPQ}%
\end{align}
where $P_{\wp\lbrack111]}$ is the payoff to $\wp$ when all three players play
$p=0$ i.e. the strategy $[1]$. Similarly $P_{\wp\lbrack221]}$ is coalition
payoff when coalition players play $p=1$ and the player in $\Re-\wp$ plays
$p=0$. Now from eq. (\ref{Qpayoff}) we get%

\begin{align}
P_{\wp\lbrack111]} &  =2P(0,0,0)=-4(\left|  c_{211}\right|  ^{2}+\left|
c_{122}\right|  ^{2})\nonumber\\
P_{\wp\lbrack112]} &  =2P(0,0,1)=2(\left|  c_{111}\right|  ^{2}+\left|
c_{222}\right|  ^{2}+\left|  c_{211}\right|  ^{2}+\left|  c_{122}\right|
^{2})\nonumber\\
P_{\wp\lbrack221]} &  =2P(1,1,0)=2(\left|  c_{111}\right|  ^{2}+\left|
c_{222}\right|  ^{2}+\left|  c_{211}\right|  ^{2}+\left|  c_{122}\right|
^{2})\nonumber\\
P_{\wp\lbrack222]} &  =2P(1,1,1)=-4(\left|  c_{211}\right|  ^{2}+\left|
c_{122}\right|  ^{2})
\end{align}
Therefore from eq. (\ref{cltPQ})%

\begin{align}
P_{\wp}  &  =-4(\left|  c_{211}\right|  ^{2}+\left|  c_{122}\right|
^{2})\left\{  lm+(1-l)(1-m)\right\}  +\nonumber\\
&  2(\left|  c_{111}\right|  ^{2}+\left|  c_{222}\right|  ^{2}+\left|
c_{211}\right|  ^{2}+\left|  c_{122}\right|  ^{2})\left\{
l(1-m)+(1-l)m\right\}
\end{align}
To find the value of coalition $\upsilon(\Gamma_{\wp})$ in the quantum game we
find $\frac{\partial P_{\wp}}{\partial m}$ and equate it to zero i.e. $P_{\wp
}$ is such a payoff to $\wp$ that the player in $\Re-\wp$ cannot change it by
changing his/her strategy given in eq. (\ref{lftQ}). It gives, interestingly,
$l=\frac{1}{2}$ and the classical optimal strategy of the coalition $\frac
{1}{2}\left[  11\right]  +\frac{1}{2}\left[  22\right]  $ becomes optimal in
the quantum game as well. In the quantum game the coalition then secures
following payoff, also termed as the value of the coalition%

\begin{equation}
\upsilon(\Gamma_{\wp})=(\left|  c_{111}\right|  ^{2}+\left|  c_{222}\right|
^{2})-(\left|  c_{211}\right|  ^{2}+\left|  c_{122}\right|  ^{2})
\end{equation}
Similarly we get the value of coalition for $\Re-\wp$ as%

\begin{equation}
\upsilon(\Gamma_{\Re-\wp})=-\left\{  \left|  c_{111}\right|  ^{2}+\left|
c_{222}\right|  ^{2}+\left|  c_{211}\right|  ^{2}+\left|  c_{122}\right|
^{2}\right\}
\end{equation}
Note that these values reduce to their classical counterparts of eq.
(\ref{VcltC}) when the initial quantum state becomes unentangled and is given
by $\left|  \psi_{in}\right\rangle =\left|  111\right\rangle $. Classical form
of the coalition game is, therefore, a subset of its quantum version. Suppose
the arbiter now has at his disposal a quantum state $\left|  \psi
_{in}\right\rangle =c_{111}\left|  111\right\rangle +c_{222}\left|
222\right\rangle +c_{211}\left|  211\right\rangle +c_{122}\left|
122\right\rangle $ such that $(\left|  c_{211}\right|  ^{2}+\left|
c_{122}\right|  ^{2})>(\left|  c_{111}\right|  ^{2}+\left|  c_{222}\right|
^{2})$. In this case $\upsilon(\Gamma_{\wp})$ becomes a negative quantity and
$\upsilon(\Gamma_{\Re-\wp})=-1$ because of the normalization given in eq.
(\ref{instate}). A more interesting case is when the arbiter has the state
$\left|  \psi_{in}\right\rangle =c_{211}\left|  211\right\rangle
+c_{122}\left|  122\right\rangle $ at his disposal. Because now both
$\upsilon(\Gamma_{\wp})$ and $\upsilon(\Gamma_{\Re-\wp})$ are $-1$ and the
players are left with no motivation to form a coalition. A quantum version of
this cooperative game, therefore, exists in which players are deprived of
motivation to form a coalition.

The payoff to a $p$ player against $q$, $r$ players in classical mixed
strategy game can be obtained from eq. (\ref{Qpayoff}) by taking $\left|
c_{111}\right|  ^{2}=1$. It gives%

\begin{equation}
P(p,q,r)=-4rq-2p+2pr+2pq+r+q
\end{equation}
Note that $P(p,q,r)+P(q,p,r)+P(r,p,q)=0$ and classical mixed strategy game is
zero-sum and $P(\wp)+P(\Re-\wp)=0$. The quantum version of this game is not
zero-sum always because from eq. (\ref{Qpayoff}) we have
\begin{equation}
P(p,q,r)+P(q,p,r)+P(r,p,q)=8(p+q+r-pq-qr-rp)-6
\end{equation}
and the quantum game becomes zero-sum only when $p+q+r-pq-qr-rp=\frac{3}{4}$.

\section{Discussion and conclusion}

There may appear several guises in which the players can cooperate in a game.
One possibility is that they are able to communicate and, hence, able to
correlate their strategies. In certain situations players can make binding
commitments before or during the play of a game. Even in the post-play
behavior the commitments can make players to redistribute their final payoffs.
The two-player games are different from multi-player games in an important
aspect. In two-player games the question before the players is whether to
cooperate or not. In multi-player case the players are faced with a more
difficult task. Each player has to decide which coalition to join. There is
also certain uncertainty that the player faces about the extent to which
players outside his coalition may coordinate their actions. Analysis of
cooperative games isolating coalition considerations instead of studying
elaborate strategic structures has drawn more attention. Recent exciting
developments in quantum game theory provide a motivation to see how forming a
coalition and its associated advantages can be influenced in already proposed
quantum versions of these cooperative games. To study this we selected an
interesting but simple cooperative game as well as a recently proposed scheme
telling how to play a quantum game. We allowed the players in the quantum
version of the game to form a coalition similar to the classical game. The
underlying assumption in this approach is that because the arbiter,
responsible for providing three qubit pure quantum initial states to be later
unitarily manipulated by the players, can forward a quantum state that
correspond to the classical game, therefore, other games corresponding to
different initial pure quantum states are quantum forms of the classical game.
This assumption, for example, reduces the problem of finding a quantum version
of the classical coalition game we considered, with an interesting property
that the advantage of making a coalition is lost, to finding some pure initial
quantum states. We showed that such quantum states can be found and,
therefore, there are quantum versions of the three-player coalition game where
the motivation for coalition formation is lost.

In conclusion, we considered a symmetric cooperative game played by three
players in classical and quantum forms. In classical form of this game, which
is also embedded in the quantum form, forming a coalition gives advantage to
players and players are motivated to do so. In quantum form of the game,
however, an initial quantum state can be prepared by the arbiter such that
coalition forming is of no advantage. The interesting function in these
situations i.e. `value of coalition' is greater for coalition then for player
outside; when the game is played classically. These values become same in a
quantum form of the game and motivation to form a coalition is lost. There is,
nevertheless, an essential difference between the two forms of the game i.e.
classical game is zero-sum but its quantum version is not.

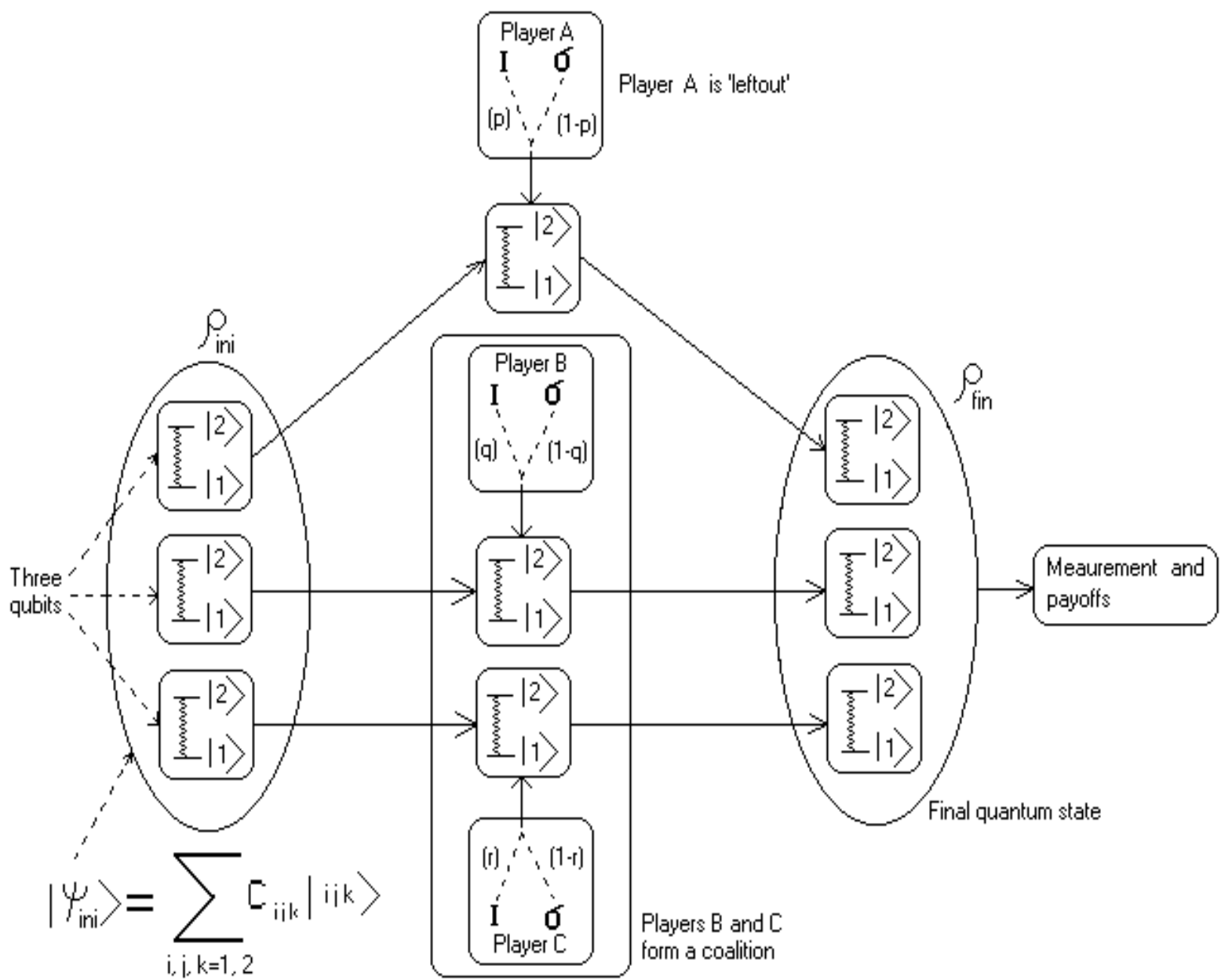
\end{document}